# Gravitation law and source model in the anisotropic geometrodynamics

Sergey Siparov

*State University of Civil Aviation, 38 Pilotov str., St-Petersburg, 196210;*
*Research Institute for Hyper Complex Systems in Geometry and Physics, 3 bg 1 Zavodskoy pr.,*
*Fryazino, Moscow region, 141190;*
*Russian Federation*

**Abstract.** The GRT modification taking into account the dependence of metric on the velocities of the sources is built. It is shown that this dependence follows from the equivalence principle and from the inseparability of the field equations and geodesics equations. As it is known, the latter are the conditions of the field equations solvability, and their form coincides with Newtonian one only in the lowest approximation. The obtained modification provides the explanation for the flat character of the rotation curves of spiral galaxies, for Tully-Fisher law, for some specific features of globular clusters behavior and for the essential excess of the observable gravitational lens effect over the predicted one. Neither dark matter nor arbitrary change of dynamics equations appeared to be needed. Important cosmological consequences are obtained.

## 1. INTRODUCTION

The flat character of the rotation curves (RC) of spiral galaxies is the most pronounced problem of the known GRT problems on the galactic scale. This is a simple, not small and statistically verified effect which doesn't find explanation not only in the GRT but in Newton gravity as well. It is to the fact that the orbital velocities of stars in the spiral galaxies don't tend to zero with the distance from the center but approach constant values of the order $10^5$m/s for all galaxies. In [1-3] there were given brief reviews and an analysis of the attempts to solve this problem, also there was suggested a new approach (anisotropic geometrodynamics or AGD) which makes it possible to overcome not only it but also the other ones known from observations and considered in [4] as necessary conditions to be sufficed by any modification of the gravitation theory. In Sections 2 and 3 the discussion of the ideas lying in the foundation of this approach is given.

Contrary to Newton theory in which massive bodies interacted instantly at large distances in the absolute and empty Euclidean space, in GRT the situation is radically different. Not only space and time became an inseparable space-time and the interaction transfer velocity became finite, but there appeared to be no bodies. Instead of them, there were the singularities of the solutions of field equations (Einstein equations) in the curved Riemannian space whose metric appeared to depend on all the four coordinates. The last means that the locations of the singularities change with time. Interpreting these singularities as physical bodies one can say together with Einstein that "the space tells the bodies how to move, the bodies tell the space how to curve". This interrelation is the fundament of GRT. In [5] Fock underlined that the key point of the GRT is the fact that the geodesics equation (or the equation of motion) is not an independent equation. It is the condition of solvability of the field equation. Therefore, while Newton theory in case a problem arises permitted to suggest some corrections to the properties of bodies, or of their interaction, or of time, or of space, and the dynamics equation was factually the definition of force and could be also changed independently – in the GRT all this is impossible, because all these notions are interlinked into one single whole. This wholeness is the undisputable achievement of the GRT and any succeeding theory may reject it only on the very solid background, more solid than the explanation of separate observations.

## 2. MODERN PROBLEMS OF THE GRT AND THE ATTEMPTS TO OVERCOME THEM

In order to overcome the problems stemming from the astrophysical observations, the following ideas are now used. The first and the most popular idea suggests to increase the right hand side of the known field equations that very well describe the situation with one point mass (Solar system), and to increase it until the obtained solution "covers" the observed effect. Physically, this corresponds to the introduction of dark matter whose mass appears to be 3-4 times larger than the mass of the visible ("luminous") baryonic matter. F.Zwicky [6] was the first to introduce this idea, then it was developed with regard to the flat RC and later it was used to interpret some modern observations [7]. The probability to verify this idea in the direct measurements does not seem too high, because of the several logical discrepancies. The very method of the dark matter mass evaluation is an a posteriori one, besides, it could be done not for every galaxy and, finally, there is an empirically grounded relation between the luminous matter (galaxy luminosity related to the luminous mass) and the observed stars' orbital velocity (Tully-Fisher law) which doesn't need dark matter.

The second idea suggested by Milgrom in the so called MOND [8] developed particularly to explain flat RC is to modify the dynamics equations or, equivalently, to modify Newton gravity law. Since the intention to fit the experiment was proclaimed as the initial one, many observational data (though not all [4]) can be described in a satisfactory way. But both the arbitrary choice of the correction term in the dynamics equations and the introduction of the new empirical world constant (measured in the acceleration units) make this theory rather applied one than scientific. The use of the dynamics equation as independent one and its arbitrary change break the self-consistency of the GRT mentioned above, this is what Fock warned of in [5]. The attempt [9] to make MOND a covariant theory lead only to the introduction of an additional scalar field, and this puts it in a row with Brans-Dicke type theories [10] but doesn't make it preferential one.

With the appearance of the observations that were interpreted with the help of the dark energy, the gravitation theory entered the crisis, because the most successful theory describing the Universe, i.e. the GRT, does it only for the 4% of the Universe energy, while the rest 96% fall upon its dark components inaccessible for the direct measurements. In order to realize the causes of this crisis, let us regard the traditional approaches of that part of the GRT which is considered to be successful. There are two of them. The first one is the approach of Einstein himself [13, 14] and of Fock [5] who both tried to construct the consecutive approximate solutions of the field equations and stopped on the second approximations in view of the technical complications. First Einstein obtained Newton gravity in [13] and then Einstein and Fock simultaneously and independently obtained the next approximations of the similar form that for the point far away from the system of bodies give the following gravitation potential [5]

$$U = \frac{G}{r}(M + \frac{E}{c^2}), \qquad (2.1)$$

where $M$ is the total mass of the bodies, and $E$ is a sum of kinetic and potential energies of the system's bodies. Formula (2.1) obtained from the field equations is in accord with the idea of mass and energy equivalence stated in the SRT.

The second approach was first used by Schwarzschild [15], and then by Friedman [16] and many others in order to investigate various concrete metrics that had believable grounds. Their advantageous features were the aesthetic appeal of the exact solutions and the vast possibilities for various interpretations – especially after Hubble's observations. But there were also common drawbacks that didn't draw enough attention due to the theoretic successes. They were the following: first, the metrics guessing was never accompanied by the geodesics reconstruction that is by the reconstruction of the equations of motion; second, they were never tested directly in experiment (the experiment technique was insufficient), and only indirect observations were used. When the testing was performed, the flat RCs for the spiral galaxies disaccording with the theoretical predictions were obtained. In view of the above mentioned circumstances pointed out by Fock, it does not seem surprising. Similar considerations dealing with the dark energy introduction are given in Section 6.2.

## 3. EQUIVALENCE PRINCIPLE, ANISOTROPY AND THEORY UNITY

In order to outline the way to overcome the problems, let us discuss some important ideas underestimated during the GRT development.

Newton gravity law based on Kepler law dealing with the planets motion in the Solar system didn't contradict the concept of the absolute space in which the Sun and planets were located. One could relate an absolute (inertial) reference frame to this space, while the conception of the non-inertial frames was first attributed to some laboratory setting and later to the Earth rotating around its axis. But for such systems as galaxies consisting of very many gravitation sources this approach is impossible. According to the equivalence principle fundamental for the GRT, there is no experimental possibility to distinguish between the gravitation forces and the inertial forces. Therefore, it is clear that we will never be able to find out from observations what kind of reference frame from the point of view of its own motion we use when observing phenomena on the galactic scale. Apart from the inertial forces related to the rectilinear acceleration of the frame (it is these forces that are usually meant when relating to the gravitation forces), there are also the inertial forces depending on the velocity of the body and the velocity of the frame, for example, Coriolis force. Then according to the equivalence principle, there are no grounds to beleive that the gravitation forces do not depend on velocities – just because as long as we are present inside this system of moving bodies and take part in the motion, there is no experimental way to distinguish them from the inertial forces. The transfer to infinity where the Coriolis type forces become also infinite needs an adequate interpretation when it comes to the experimental check up. It is this that makes the situation in a galaxy different from the situation in a planetary system in which the phenomena taking place in the vicinities of the star, of the planet and in the laboratory can be easily separated according to their scale, and the frame with the definite motion properties with regard to these phenomena can be chosen. Thus, the idea to accept the gravitation force dependence on the velocities of the probe body and of the surrounding system of sources seems natural.

As it is known, any theory that consistently unifies the inverse square law for the forces and Lorentz invariance has to contain the field originated from the currents (see the discussion in [2,3]). In electrodynamics where the interaction constant $q$ is the electric charge, this field is magnetic vector field, and it produces the Lorentz force which depends on the charges' velocities and enters the right hand side of the geodesics equation. The expression for it is

$$\vec{F}_L = q\vec{E} + q[\vec{v}, \vec{B}] \qquad (3.1)$$

In the approach known as gravitoelectromagnetism (GEM) [17], the gravitation is described with the help of scalar and vector potentials analogously to electrodynamics. Although the corrections to the classical GRT calculated with regard to the massive body rotation do not contradict the measurements, this analogy is too straightforward and can not be applied to gravitation directly. GEM theory copies electrodynamics formally, but in the latter the velocity dependent Lorentz force is produced by the additional (external) field and corresponding sources (electric charges). In the GRT, the ineradicable gravitation force can not be attributed to an external (additional) field, similarly, no part of it can be arbitrarily regarded as an additional field as in GEM. Being the cause of the geometrical curvature according to the equivalence principle, the gravitation field must be considered directly in the metric. The same considerations can be related to an attempt to apply the theory of retarded (or Lienard-Wiechert) potentials to gravitation.

Turning back to the velocity dependence of gravitation forces, we conclude that the metric corresponding to the potentials of the gravitation forces must include the vector fields related to the velocities of the sources and of the probe particle. The velocity dependent metric characterizes the anisotropic space whose geometry is no longer Riemannian one.

Attempting to construct or modify the theory, one must pay attention to the preservation of its unity achieved in the GRT, i.e. to preserve the interrelation between the geodesics equations and the field equations in the process of solution. Notice, that we are interested not only in a field equation solution (metric) as it is, but we want to substitute it into the equation of motion and compare the result with observations.

As it was mentioned, the geodesics equation is not independent, and the motion equation in the Riemannian space takes Newton form only in the lowest approximation of the metric expansion in the vicinity of the Minkowski metric. Even in the next approximation in the Riemannian space, the corrections to it must be considered [5]. But they can't be inserted into the dynamics equations in an arbitrary way without considering the field equations. This is also true for the space which we are going to consider and in which the geodesics equations are supposed to take another form.

## 4. METRIC, MOTION EQUATIONS AND GRAVITATION LAW

Let us obtain the general form of the motion equation corresponding to the solution of the field equation. We also take the metric in general form $\widetilde{g}_{ij}(x,u(x),y)$ and consider it anisotropic with regard to the above said. As usual, we consider the weak field and take the linear part of the solution. Let us present the metric as [2]

$$\widetilde{g}_{ij}(x,u(x),y) \equiv g_{ij}(x,y) = \gamma_{ij} + \varepsilon_{ij}(x,y) \qquad (4.1)$$

where $\gamma_{ij}$ is Minkowski metric; $\varepsilon_{ij}(x, y)$ is a small anisotropic perturbation; $y$ belongs to the tangent space, and along the probe trajectory $x^i = x^i(s)$ there is always $y^i = \dfrac{dx^i}{ds}$; finally, $u(x)$ is a vector field describing the self-consistent motion of the sources and producing the anisotropy. The geometry for which the metric is given by eq. (4.1) is called the generalized Lagrange geometry.

In order to obtain the generalized geodesics corresponding to our approximation, let us variate the Lagrangian $L = (\gamma_{hl} + \varepsilon_{hl}(x,y))y^h y^l$ as in [18]. Then the generalized geodesic takes the form

$$\frac{dy^i}{ds} + (\Gamma^i{}_{lk} + \frac{1}{2}\gamma^{it}\frac{\partial^2 \varepsilon_{kl}}{\partial x^j \partial y^t}y^j)y^k y^l = 0 \qquad (4.2)$$

where $\Gamma^i{}_{jk} = \frac{1}{2}\gamma^{ih}(\frac{\partial \varepsilon_{hj}}{\partial x^k} + \frac{\partial \varepsilon_{hk}}{\partial x^j} - \frac{\partial \varepsilon_{jk}}{\partial x^h})$ is a Christoffel symbol depending on $y$. Thus, in order to obtain the motion equation for the exact solution of the field equation in the weak field limit, one should use eq. (4.2) and not the usual geodesics equation $\frac{dy^i}{ds} + \Gamma^i{}_{lk} y^l y^k = 0$ valid in the Riemannian space. Taking the same well-known assumptions and repeating the same calculations as were performed by Einstein in [13], and then extracting the anti-symmetric part of the auxiliary tensor introduced in [2,3], one obtains the space domain part of the motion equation obtained from the geodesics equation (4.2)

$$\frac{d\vec{v}}{dt} = \frac{c^2}{2}\left\{-\nabla\varepsilon_{11} + [\vec{v}, rot\frac{\partial \varepsilon_{11}}{\partial \vec{v}}] + \nabla(\vec{v}, \frac{\partial \varepsilon_{11}}{\partial \vec{v}})\right\} \quad (4.3)$$

Here $\varepsilon_{11}$ is the only component of the metric tensor that remains in the motion equation when the weak field and the small velocities assumptions are used. Using eq. (4.3) as the motion equation, we get the expression for the generalized gravitation force [2, 3]

$$\vec{F}^{(g)} = \frac{mc^2}{2}\left\{-\nabla\varepsilon_{11} + [\vec{v}, rot\frac{\partial \varepsilon_{11}}{\partial \vec{v}}] + \nabla(\vec{v}, \frac{\partial \varepsilon_{11}}{\partial \vec{v}})\right\} \quad (4.4)$$

As in [14], we obtained the mechanics equation out of the field equations only, no special choice of momentum-energy tensor and no additional a priori suggestions about the form of the geodesics equations were made. Let us discuss the physical sense of the components of the gravitation force present in eq. (4.4).

The first term is related to the expression of the Newtonian force $F^{(g)}{}_N$ acting on a particle with mass $m$. For a single stationary spherically symmetric gravitation source with mass $M$, let us introduce $\varepsilon_{11}(x) = \frac{r_S}{r}$ where $r_S = \frac{2GM}{c^2}$ is the so called Schwarzschild radius and get the known Newton law for gravitation.

If we introduce

$$\vec{\Omega}(x) = \frac{c^2}{4} rot\frac{\partial \varepsilon_{11}}{\partial \vec{v}}, \quad (4.5)$$

then the second term in eq. (4.4) can be recognized as an analogue of "Coriolis force" $F^{(g)}{}_C = 2m[\vec{v}, \vec{\Omega}(x)]$ which is intrinsic to the non-inertial (uniformly rotating) reference frame. It is proportional to the velocity, $\vec{v}$ of the probe particle and also depends on the frame motion[1]. Now this force is related to the motion of the gravitation sources. Notice, that the actions produced by the gravitation force component $F^{(g)}{}_C$ could be attraction, repulsion and tangent action depending on the angle between $\vec{v}$ and $\vec{\Omega}(x)$. If in a certain region vector $\vec{\Omega}(x)$ preserves constant direction, then the component of the body's velocity, $\vec{v}$ which is parallel to $\vec{\Omega}$ doesn't suffer the action of the second term in eq. (4.4).

Alongside with the flat RC, there is another known observational paradox [4]. Contrary to the orbital motion of stars on the periphery of a spiral galaxy, the motion of the globular clusters in the plane orthogonal to the galaxy plane is in agreement with the calculations based on Kepler law. The idea of the additional component of the gravitational force $F^{(g)}{}_C$ acting on the stars moving in the galaxy plane but not acting on the clusters moving orthogonally to it explains this paradox.

---

[1] If we designate $\frac{c^2}{2} rot\frac{\partial \varepsilon_{11}}{\partial \vec{v}} = \vec{B}^{(g)}(x) = 2rot\vec{u}$, then we can speak of the gravitational analogue of Lorentz force. This could be convenient from the point of view of calculations and will be used later.

The third term in eq. (4.4) presents an additional force of attraction or repulsion acting on a moving particle when the system of gravitation sources extends radially (explosion) or contracts radially (collapse). This is interesting from the point of view of describing and calculation of the corresponding astrophysical processes.

Let us transform the expression for the gravitation force obtained for the space with metric (4.1) depending on the velocities of the distributed sources. In order to get it in the form more convenient for the experimental check up, let us fix a certain reference body. Now introduce the notation

$$\vec{u} \equiv \frac{c^2}{4} \frac{\partial \varepsilon_{11}}{\partial \vec{v}} \equiv [\vec{\Omega}, \vec{r}], \qquad (4.6)$$

where $r$ is a radius vector of the point, then unite the first and the third components in eq. (4.4) under the common gradient sign

$$\vec{F}^{(g)} = \frac{mc^2}{2} \nabla \{-\varepsilon_{11} + \frac{2}{c^2} \cdot 2(\vec{u}, \vec{v})\} + 2m[\vec{v}, rot\vec{u}] \qquad (4.7)$$

and introduce the scalar potential $([\vec{v}, rot\vec{u}], \vec{r})$ for which the second term in eq. (4.4) will be also a gradient. Now use the relation $([\vec{v}, rot\vec{u}], \vec{r}) = (\vec{v}, [rot\vec{u}, \vec{r}]) = (\vec{v}, \vec{u})$ and finally get the expression

$$\vec{F}^{(g)} = \frac{mc^2}{2} \nabla \{-\varepsilon_{11} + \frac{2}{c^2} \cdot 4(\vec{u}, \vec{v})\}. \qquad (4.8)$$

This is the gravitation law obtained from the dynamics equation (geodesics equation) which is the condition of solvability of the field equations in the general case. When calculating, one should consider all the bodies taking part in motion. Alongside with the common term related to Newton potential, in the new equation there is an anisotropic scalar potential whose role starts to be important at some conditions. Notice, that similarly to [14], no suggestions about the concrete form of the metric were made, and the only assumption is the usual weak field approximation.

Demanding Newton character of the gravitation law for the limit case of one body, we get

$$\vec{F}^{(g)} = \frac{mc^2}{2} \nabla \{-\sum_n \frac{r_{n,S}}{r_n} + \frac{2}{c^2} \cdot 4(\vec{u}, \vec{v})\}. \qquad (4.9)$$

Despite the formal similarity, the expression (4.9) can't be regarded as a post-Newtonian approximation for the two following reasons. First, there is no central symmetry in the problem in question. Second, the physical meaning of their difference can be seen in the fact that in a post-Newtonian approximation, the correction to Newton law starts to play when the point approaches the center, while in our case the second term becomes important when the point moves away from the origin, and the Newtonian term goes out. The calculation of $\vec{u}(x)$ is performed similarly to electrodynamics, since as it was underlined in [2, 3] both field theories have the same geometric origin. This means that the same mathematical results can be used for both cases. Applying the known magnetic induction formula for the gravitation case, one gets

$$\Omega = rot(\int \frac{j^{(m)}(r)}{|r - r_0|} dV), \qquad (4.10)$$

where $j^{(m)}(r)$ is the mass current density and $r_0$ is the observers coordinate. The concrete examples of its use will be given later.

The new circumstance is the possibility of the negative correction to Newton potential – up to the predominance of (observable) repulsion over attraction forces. But this is of no surprise, because it is related to the equivalence principle and to the identifying of the gravitational and inertial forces. The latter include, for example, the centrifugal forces. The additional attraction forces (with regard to Newton ones) that are due to the motion can be also found. But this is also in agreement with the GRT ideas that can be seen in eq. (2.1). Since we broke the GRT formalism nowhere, all the obtained expressions are, naturally, Lorentz invariant. The matrix of the coordinates ($x^i$) transformation is the same as of the corresponding derivatives ($y^i$) transformation. All the bodies including the one with the observer and the probe body move along their geodesics. Therefore, when we pass to reference frames related to other bodies, the calculated trajectories will, naturally, change their descriptions, but no new events (like extra collisions) can appear.

Let us make several remarks. First, let us compare the obtained expression for the gravitation force eq. (4.7) and the expression for the force in the non-inertial system. Taking for simplicity the uniformly rotating frame, we see that the force is equal to

$$\vec{F}_{N-I} = m\nabla(\Phi + \frac{V^2}{2}) + m[\vec{v}, rot\vec{V}], \qquad (4.11)$$

where $\Phi$ is the force potential, $v$ is the probe velocity, $\vec{V} = [\vec{\omega}, \vec{r}]$ is the linear velocity at distance $r$ from the rotation axis measured in the motionless frame. The second term in eq. (4.11) is proportional to Coriolis force; the second term in brackets under gradient is called the centrifugal potential energy. As $r$ approaches infinity, this energy approaches infinity too.

Second, let us compare the potential in formula (4.9) and Fock's result eq. (2.1). The difference is due to the fact that equation (2.1) describes the external problem and deals with the potential in the point which is far away from the system of bodies. In our case it is not so. The probe body is one of the bodies of the system, and the potential is defined in the point which the probe occupies moving inside the distribution of moving masses. The specific features of the AGD potential calculated for the point far away from the system will be regarded in Section 6.2.

## 5. MODEL OF GRAVITATION SOURCE AND ITS APPLICATIONS

In the AGD, the singularities in general case may possess more complicated properties than points, because the potential now depends not only on coordinates but also on the vector field that could have a solenoidal part. Therefore, for the physical interpretations of such singularities as the sources of the gravitation field, the simplest (basic) model of the source will be not the point (spherically symmetric body) as in the GRT, but a system like a "center plus current" (CPC), i.e. the central body surrounded by the effective mass current $J^{(m)}$. The spiral galaxies are obviously the objects suitable to be described by the CPC model. In order to describe more complicated systems of moving masses the model has to be more complicated too.

Let us here limit ourselves with the spiral galaxies case. For such galaxies as M-104 (Sombrero) or NGC-742 (Fig. 1a, b) that have rings, the CPC model can be applied directly. For other spiral galaxies – with more pronounced arm structure – one can introduce the effective values of the contour radius, $R_{eff}$, constant angular velocity, $\Omega_{eff}$, and linear velocity of the mass current density along the contour, $V_{eff} = \Omega_{eff} R_{eff}$. This can be done, for example, in the following way

$$I_{eff} = \sum I_n \equiv MR_{eff}^2 \Rightarrow R_{eff}^2 = \sqrt{\frac{I_{eff}}{M}} \qquad (5.1)$$

where $I_{eff}$ is the moment of inertia of the system with the total mass $M$. The effective angular velocity $\Omega_{eff}$ can be obtained from the condition $I_{eff}\Omega_{eff} \equiv L_{eff} = \sum L_n$, where $L_n$ is the angular momentum of the system components. Thus, we get

$$\Omega_{eff} = \frac{L_{eff}}{I_{eff}} . \qquad (5.2)$$

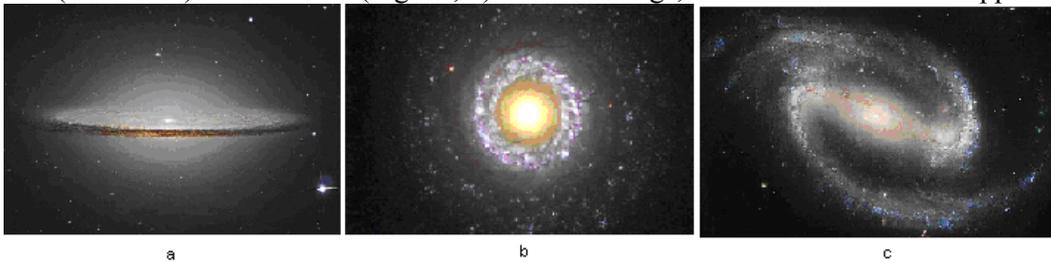

**FIGURE 1**. Spiral galaxies**:** a – M-104, b – NGC-7742, c – NGC-1300. (The images are obtained by the Hubble telescope)

These parameters can be obtained for any concrete galaxy from the astronomical observations. The relation between the phenomenological value $\Omega_{eff}$ and the general expression $\Omega(x)$ introduced earlier is obvious.

Let us find the conditions for which the regular GRT model (point) is not enough to describe the physical system, and one has to use the CPC model of the AGD. These conditions take place when both terms in the eq. (4.9) start to play comparable roles. Let us take that the probe body, i.e. one of the stars on the periphery of the spiral galaxy, moves in the same way as the bodies that we consider to belong to the circular current, so that $u \sim v$. Then from the formula (4.9) it follows that the required condition for the distance between the star and the center of galaxy and for the star velocity is

$$rv^2 \sim \frac{1}{4}GM . \qquad (5.3)$$

For a galaxy with mass $M$, with the radius of its visible disk $r$, and for the velocity $v$ of the stars moving at its periphery, this condition corresponds to the values observed in astrophysics on galactic scale. Therefore, the discrepancy between Newton-Einstein-Schwarzschild predictions and the observations is not surprising.

Thanks to the common geometric origin of "Maxwell equations" in electromagnetism and gravitation discussed in [2, 3], the mathematical results obtained in electrodynamics can be used to calculate the motion of the gravitational systems. One can notice that the gravitational CPC model is analogous to an electromagnetic model that consists of a circular electric current and a charge in the center of it.

## 5.1 Flat Rotation Curves

Let us for convenience first speak in terms of the electromagnetic version of the CPC model, i.e. regard a positive charge surrounded by a circular current $J$, and an electron circling around in the plane of the contour and close to it. Strictly speaking, an electron in such a system can not be in a finite motion and has either to fly away or to fall on the center. But the number of rotations performed before that could be large enough. The value of $B_z(r)$ component of the magnetic induction produced by the contour with radius, $R_{eff}$, can be found with the help of Bio-Savart law and according to [19] with $c = 1$ is equal to

$$B_z(r) = J \frac{2}{\sqrt{(R_{eff}+r)^2+z^2}} [K + \frac{R_{eff}^2 - r^2 - z^2}{(R_{eff}-r)^2+z^2} E]$$

$$K = \int_0^{\pi/2} \frac{d\theta}{\sqrt{1-k^2\sin^2\theta}}; E = \int_0^{\pi/2}\sqrt{1-k^2\sin^2\theta}d\theta \qquad (5.4)$$

$$k^2 = \frac{4R_{eff} r}{(R_{eff}+r)^2+z^2}$$

where $K$ and $E$ are elliptic integrals. Introducing the notation, $b = r/R_{eff}$ and taking $z = 0$, one gets

$$B_z(r) = J \frac{2}{R_{eff}(1+b)}[K + \frac{1-b^2}{(1-b)^2}E] \qquad (5.5)$$

The internal region close to the charge corresponds to $b << 1$ and to the constant value $B_z(r) \to J/2R_{eff}$; the remote region corresponds to $b >> 1$ and to $B_z(r) \to 0$; the intermediate region to which the contour belongs and in which the electron moves corresponds to $b = O(1)$ and

$$B_z(r) \sim J/r \qquad (5.6)$$

The centrifugal force, $m\frac{v_{orb}^2}{r}$ acting on electron is equal to the sum of Coulomb attraction, $F_{Cl} = qC_1/r^2$ produced by the central charge and of Lorentz force $F_L = qv_{orb}B_z(r)$

$$\frac{mv_{orb}^2}{r} = \frac{qC_1}{r^2} \pm qv_{orb}B_z(r) \qquad (5.7)$$

Passing to the gravitational version of the CPC, we change the electric charge to gravitational one, i.e. $q = m_g$, then use the equivalence principle, i.e. $m_g = m$, and finally consider that the region corresponding to the periphery of a galaxy is the intermediate one, i.e. $B^{(g)}_z(r) \sim J/r$. For a spiral galaxy its effective radius does not differ much from the radius of the visible disk where the measurements of the stars orbital velocities are possible. Therefore, we can neglect the change of the character of $B_z(r)$ decrease with $r$. Then we designate $J \equiv C_2$ and get the dynamics equation in the form

$$v_{orb}^2 = \frac{C_1}{r} \pm v_{orb}C_2 \qquad (5.8)$$

where $C_1$ and $C_2$ are constants characterizing the system, and the sign corresponds to the current direction and to the location of the probe inside or outside the contour. The smaller root of the quadratic equation (5.8) gives the expression $v_{orb} = \frac{C_2}{2}(1-\sqrt{1\pm\frac{4C_1}{rC_2^2}})$, corresponding to Newton law in which the sign corresponds to the direction of the probe motion. Neglecting the small term under the square root in the larger root of the quadratic equation (5.8), i.e. in the expression $v_{orb} = \frac{C_2}{2}(1+\sqrt{1\pm\frac{4C_1}{rC_2^2}})$, one gets

$$v_{orb} \sim C_2, \qquad (5.9)$$

This corresponds to the flat RC at the periphery and neither needs dark matter, nor presumes an arbitrary change of the dynamics equation.

## 5.2 Tully-Fisher Law

Let us evaluate $C_2(R_{eff}) = J^{(m)}(R_{eff})$. The mass current can be given as $J^{(m)}(R_{eff}) \sim M/T$, where the spiral galaxy mass, $M$ is proportional to its area, i.e. to $R_{eff}^2$, and the period can be evaluated according to Kepler law as $T \sim R_{eff}^{3/2}$. This will give $J^{(m)}(R_{eff}) \sim \sqrt{R_{eff}}$. Since the integral luminosity, $L_{lum}$ is also proportional to the area of the galactic disk, we get $R_{eff} \sim \sqrt{L_{lum}}$. Therefore, $J^{(m)}(R_{eff}) \sim \sqrt{R_{eff}} \sim L_{lum}^{1/4}$ and finally

$$v_{orb} \sim L_{lum}^{1/4} \tag{5.9}$$

which is in accord with the empirical Tully-Fisher law. We can notice that the exponent ¼ corresponds to the suggestion of the constant mass to luminosity ratio which is used as one of the main propositions when measuring distances in cosmology.

### 5.3 Logarithmic Potential

Being applied to a spiral galaxy whose matter is distributed mostly in the plane, the obtained results have transparent physical sense related to the field equations. Let us regard expression (4.6). From the measurement units' considerations, it follows that $\varepsilon_{11} \sim \ln(v/V_{eff})$. If velocity, $v$ in this expression is the velocity of a star which is an element of the mass current, then $v \sim u$, and according to the used notations, $u = V_{eff} r / R_{eff}$. As it is known, the logarithmic potential, $\varepsilon_{11} \sim \ln(r/R_{eff})$ is the solution of the two-dimensional Poisson equation which can be obtained here similarly to [13] from the field equation in the same approximation. In order to do this, one has to neglect the motion in the direction orthogonal to the galaxy plane in Einstein tensor, and in the energy-momentum tensor account for the fact that the mass belongs not only to the center but also to the ring. Therefore, the expression for the component of the metric tensor will contain the solution of Poisson equation not only for the three-dimensional problem, $\varepsilon_{11} \sim 1/r$ (Newton potential), but also for the two-dimensional problem, $\varepsilon_{11} \sim \ln r$. Such logarithmic potential was suggested in [20] from heuristic considerations in order to describe the gravitation in the spiral galaxy. There the authors were searching for the way to describe the observations without the use of dark matter, they expressed their hope that such potential could be obtained from the relativistic considerations. The same potential was discussed in [21] where it was introduced to describe a certain (2+1)-dimensional scalar field in the cosmological domain wall whose plane contains the planes of spiral galaxies. It can be seen that in the AGD the logarithm for a spiral galaxy naturally appears from the anisotropic metric and does not need any additional field to be introduced. When the moving masses are distributed in a more complicated way, the potentials will obtain a related form in accord with eq. (4.6).

## 6. CLASSICAL GRT TESTS ON THE GALAXY SCALE

The important achievement of the GRT was the prediction and subsequent observation of the so called classical tests: orbit precession, light beam bending and gravitational red shift. Let us use the CPC model in order to find out what do these phenomena correspond to on the galactic scale. The visual results can be obtained with the help of numerical calculations which is due mostly to the presence of the elliptic integrals in the functions' descriptions.

### 6.1 Orbit Precession

In order to describe the motion of a star in a spiral galaxy around the galaxy nucleus, let us use the CPC model and regard a particle orbiting in the contour plane. Choosing various initial conditions, one can obtain scattering, swift fall down on the center and long enough orbiting (see examples on Fig.2). If the initial conditions lead to the large enough number of particle rotations around the CPC center, then Fig.2a illustrates what could be called quasi-precession. This type of a star motion is the galactic scale AGD-analogue of the GRT orbit precession in the planetary system scale. Unfortunately, it is not so easy to verify the correspondence between the calculated and observed motions on the galactic scale. The idea of the (elliptical) stars' orbits precession was used in [22] in order to develop the density wave theory of the arms origination. The trajectory given on Fig.2b explains why the globular clusters can be frequently found in the vicinity of the galaxy center and not on the periphery. According to Kepler's idea of the elliptical orbits which is the basis of Newton gravitation theory, it is the periphery where they should spend the major part of their lifetime. But now we see that if the plane of a cluster orbit is close to the galaxy plane, we cannot neglect the influence of the mass current which can produce the situation given on Fig.2b, i.e. hold the cluster close to the center. The explanation of this feature of the globular clusters motion was also mentioned in [4] as one of the demands to be sufficed by any modification of gravitation theory.

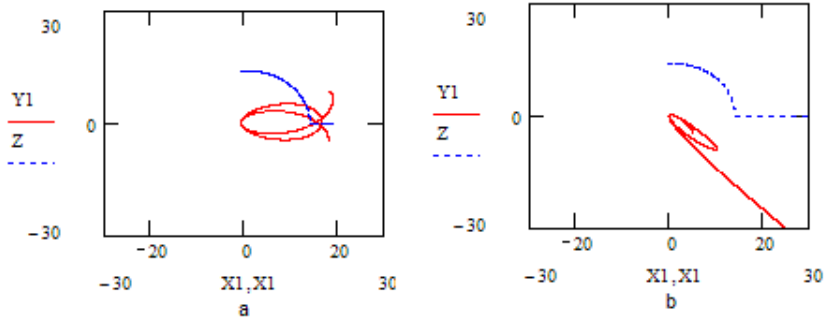

**FIGURE 2.** Trajectories obtained with the help of the AGD basis model of the source: a – quasi-precession; b – globular clusters' motion disagreeing Kepler law

It should be mentioned that the numerical modeling of the CPC system provides a lot of patterns resembling the observed ones. This makes it possible to suggest new hypotheses of the origination of arms and also of bars characteristic for 2/3 of spiral galaxies, the origination of bars being unclear up to now.

## 6.2 Light Bending

Fig.3 contains the comparison of the scattering of the probe body on a Coulomb center and on a CPC as suggested by the AGD. Since the expression (4.1) for the metric which has to be used for the calculation of the light bending angle now depends on the proper velocities of the sources, these trajectories can be related to the light beams trajectories. The comparison of Fig.3a and Fig.3b explains why the observed values of the light bending by the gravitational lenses can be essentially (several times) larger than those predicted by the GRT, and no dark matter is needed. Moreover, Fig.4 illustrates a probe body scattering on a CPC along the trajectory which bends twice in the opposite directions – this is caused by the change of the direction of mass current present on Fig.3b. From the point of view of light refraction, it means that the AGD predicts the existence of the diffracting gravitational lenses that diminish the angular sizes of the objects located behind them. It means that if such a lens is located between the object and the observer on the Earth and is oriented in a due manner, then the estimation of the distance to the object could higher than it really is. If this object is an astronomical standard candle, this situation could be the reason of the misinterpretations of the last decade observations concerning type 1a supernovae presumably showing the break of Hubble linear law [23]. There were these observations that lead to the idea of the accelerated Universe expansion and to the introduction of the notion of dark energy of repulsion providing the acceleration. It caused the appearance of the new cosmological constant (of the opposite sign) in Einstein equation. Now we see that the interpretation of the SN1a observations with regard to the possible existence of diffracting gravitational lenses provides the possibility not to demand dark energy existence from the Universe, while all the theoretical results of the classical GRT remain valid in its region of applicability.

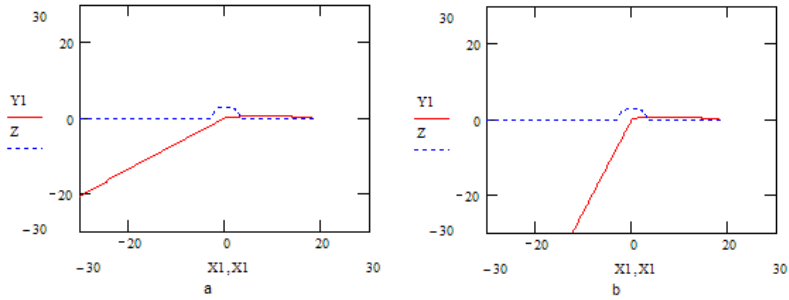

**FIGURE 3.** Trajectories of a probe body scattering: a – on the Coulomb center; b – on the CPC

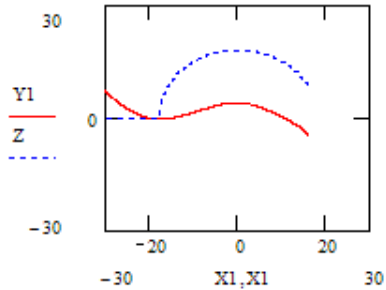

**FIGURE 4.** Trajectory of a probe body scattering on the CPC: double bending with opposite signs

Let us give the considerations mentioned in the end of Section 2. As it is known, for the first time the cosmological constant was introduced by Einstein to provide the observed stationary character of the Universe in frames of the GRT. Its physical meaning corresponded to the existence of the additional energy of attraction characteristic for the cosmological scale. Einstein considered the existence of this (directly unobservable) energy more probable than the existence of the (directly unobservable) Universe expansion presumed by Friedman solution, which Einstein considered to be mathematically correct but lacking attitude to physical reality. After Hubble's observations and their interpretation in terms of Friedman theory, Einstein concluded that the introduction of the cosmological constant was erroneous. Both situations with those constants introductions are alike from the point of view of the inability of the acknowledged theory to suffice the observations.

In frames of the AGD, the gravitational potential in eq. (4.9) contains an additional term which can have different signs depending on the object and on the conditions. This makes the introduction of an absolute constant less natural than the use of the local corrections due to the relative motion of the parts of the Universe.

## 6.3 Gravitational Red Shift

The fact that the kinematical effects are indistinguishable from the dynamical effects on the galactic scale, i.e. the dependence of metric on velocity, means that the Doppler type effect caused by the uniform motion of the frame is indistinguishable from the gravitational shift which in some cases may be not obligatory red.

When we pass to the consideration of the gravitational potential far from the observer in frames of the AGD, there appears a principal feature which characterizes the anisotropic space introduced in order to take a more detailed account for the equivalence principle and for observational possibilities. Let the frame origin (the observer) be located on one of the bodies of the system. When the point in which we are going to determine the gravitational potential approaches the infinity, the potential tends to zero according to eq. (4.9) and in a way similar to that of the law defined by eq. (2.1). The energy of the system also enters the effective mass which remains finite. Now let the body with the observer become further and further from the system, and let us defines the potential in the point close to one of the bodies of the system. Then the total kinetic energy of the system measured relative to the reference body will enter the potential. This corresponds to the situation when a system of massive bodies (or a single body) obtains an observable gravitational potential growing with the distance from the observer. This statement seems strange at the first sight, but it has a direct connection with Mach's ideas, it can be experimentally verified and it will lead to the observation of the gravitational red shift which will grow with the distance from the selected source to the observer. Actually, the observation of Hubble red shift can be interpreted as a manifestation of this very effect predicted by the AGD. This means that the red shift can be due not only to the radial expansion of the Universe, but also to the tangential motion of its parts. Notice that in accord with the expression for the gravitation potential in eq. (4.9), the first term tends to zero when the observer moves away from the system, but the second term grows as $(\vec{u},\vec{v}) \to v^2 \sim r^2$. Therefore, the time interval will be inversely proportional to the distance to the (tangentially) moving source, and this corresponds to the linear Hubble law. Data scattering and the observed anisotropy of the Hubble law could be due to the various values of $\Omega(x)$ factor standing in front of $r$.

Such interpretation has an indirect observational support presented by the measurements of the tangential velocities of the far away quasars given on Fig.5 – the velocities appeared to be surprisingly high [24].

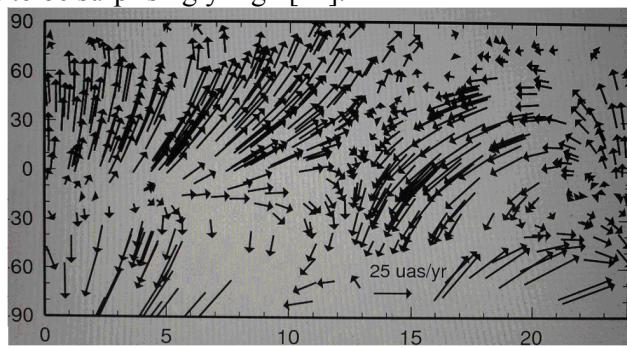

**FIGURE 5.** Observed tangential motion of quasars [24]

Thus, the phenomena known as the GRT classical tests not only correspond to the ideas of the AGD and can be calculated but are directly observed. One can see that they appear to be related just to those paradoxical astrophysical observations that have no explanation in the GRT but are natural for the AGD.

# 7. AGD SPECIFIC TEST

Despite the presence of drawbacks caused by the usual interpretations of the observations that include the notions of dark matter and dark energy, and despite the absence of them in the interpretations based on the AGD, one has to suggest an experiment which could directly show the metric tensor dependence on the moving masses velocities on the galactic scale. As it was shown in [1] in more detail, such experiment can be based on the investigation of the effect of the optic-metrical parametric resonance (OMPR) for various astrophysical systems in our galaxy. The theory of the OMPR was developed in [25], and the recently performed observations lead to the first encouraging results [26] pointing at the OMPR existence. The OMPR effect reveals when the gravitational waves emitted by the periodic source (close binary star) resonantly interact with the electromagnetic radiation of the space maser. If certain conditions of the OMPR are fulfilled, the signal can be received by the usual radio telescope. If the metric tensor has an anisotropy related to the rotation of Milky Way which is also a spiral galaxy, then the conditions of the OMPR will depend on the orientation of the "gravitational waves source – space maser" system relative to the galactic plane. This can be registered in observations and the corresponding astrophysical systems were suggested in [1]. In [18] the calculation of the anisotropy influence on the OMPR conditions was performed. The comparison with the corresponding observations would become possible when the amount of the observational data will be enough for the statistical analysis.

# 8. CONCLUSIOS AND DISCUSSION

The main conclusion of this paper is that there are solid grounds for the reinterpretation of the available observation data obtained on the galactic and higher scales. It appears to be possible to consistently get over the discrepancies between the GRT predictions and the observations without the help of the dark notions introduction or of the arbitrary change of the equations. Fortunately, this can be done without the fundamental breaking of the GRT foundations. At the same time, the cosmetic attempts focused on the metric guessing, even such famous as Schwarzschild metric or Friedman metric are inadequate. The cause of it is the indissoluble interrelation of the field equations and the geodesics equation underlined by V.A.Fock. Such circumstances as the account for the equivalence principle in its full meaning, the general properties of the Lorentz-invariant fields corresponding to the forces described by the inverse square law, geometrical origin of "Maxwell equations" simultaneously demand and permit to use the gravitation law that accounts for the velocities of the gravitating bodies. This makes it necessary to introduce the anisotropic metric and the corresponding anisotropic space. The obtained general expressions for the geodesics, for the gravitation force and potential, on the one hand, have the known GRT and Newton laws as their limits, and on the other hand – on the galactic scale – correctly describe the observations without the use of dark matter. The known GRT classical tests on the galactic scale correspond to observations – to those of them that encounter problems within the classical GRT. There also appears the theoretical possibility to do without the dark energy, though this still requires experimental check up.

The results of the AGD approach are: all the results of the classical GRT remain valid on the planetary system scale; the discrepancies between the theory and observations on the galactic scale are eliminated; the specific experiment able to reveal the influence of the anisotropy on the galactic scale is suggested, the calculations and the first encouraging observations for it are performed.

The AGD approach is suitable also for larger scales, i.e. for the galactic clusters and for the large scale structure of the Universe. But these problems – as well as many others stemming from the reinterpretations of the observations – are beyond the limits of this paper.

## 9. ACKNOWLEDGEMENTS

This work was performed in frames of the RFBR grant No. 07-01-91681-RA\_a. The author expresses his gratitude to D.Pavlov for support and to N.Voicu, S.Kokarev, N.Razumovsky and A.Kazakov for helpful discussions.

## 10. REFERENCES

[1] – S.V.Siparov. (2006), HNGP 2(6), c.155 (rus); S.Siparov. (2007), In "*Space-Time Structure: Algebra and Geometry*", p.495, Moscow; S.Siparov. (2008), Acta Mathematica APN, 24(1), p.135
[2] – S.Siparov. (2008), arXiv [gr-qc]: 0809.1817 v3
[3] – S.V.Siparov. (2008), HNGP 2(10), c.64 (rus)
[4] – A.Aguirre, C.P.Burgess, A.Friedland and D.Nolte. (2001), CQG 18, R223
[5] – V.A.Fock. (1939), JETP 9(4), c.375 (rus)
[6] – F.Zwicky. (1937), ApJ 86, p.217
[7] – D.Clowe et al. (2006), ApJ Lett. 648, L109
[8] – M.Milgrom. (1983), ApJ. 270, p.384
[9] – J.D. Bekenstein, and M.Milgrom. (1984), ApJ 286, p.7
[10] – C.Brans and R. H.Dicke. (1961), Phys. Rev. 124, p.925
[13] – A.Einstein. (1916), Ann.d.Phys. 49, p.769
[14] – A.Einstein, I.Infeld and B.Hoffmann. (1938), Ann.Math. 39 (№1), p.65
[15] – K.Schwarzschild. (1916), Sitzungsber.d.Berl.Akad. p.189
[16] – A.Friedmann. (1922), Zs.Phys., 10, p.377
[17] – M.L.Ruggiero and A.Tartaglia. (2002), arXiv: [gr-qc] 0207065v2
[18] – S.Siparov & N.Brinzei. (2008), arXiv: [gr-qc] 0806.3066; N.Brinzei, S.Siparov. (2007), ГЧГФ 4, c.41; S.Siparov, N.Brinzei. (2008) ГЧГФ 5, c.56
[19] – L.D.Landau, E.M.Lifshits. (1957), «*Electrodynamics of the Continuous Media*», Moscow (rus).
[20] – W.H.Kinney & M.Brisudova. (2000), arXiv: [astro-ph] 0006453
[21] – G.Dvali, G.Gabadadze & M.Shifman. (2001), arXiv: [astro-ph] 0102422v2
[22] – C.C.Lin & F.H.Shu. (1964), ApJ 140, p.646
[23] – A.Riess et al. (1998), Astronomical Journal 116, p.1009
[24] – D.S. MacMillan. (2003), arXiv: [astro-ph] 0309826
[25] – S.V.Siparov. (2004), Astron. & Astrophys., 416, p.815
[26] – S.V.Siparov & V.A.Samodurov. arXiv[astro-ph] : (2009) 0904.1875; S.V.Siparov, V.A.Samodurov. (2009), Comp.Opt. 33 (1), c.79 (rus)